# Effect of temperature on aging and time-temperature superposition in nonergodic Laponite suspensions

Varun Awasthi and Yogesh M Joshi*

Department of Chemical Engineering, Indian Institute of Technology Kanpur, Kanpur 208016, INDIA.

* Corresponding Author: E-Mail: joshi@iitk.ac.in

Telephone: +91 512 259 7993, Fax: +91 512 259 0104

**ABSTRACT**

We have studied the effect of temperature on aging dynamics of laponite suspensions by carrying out the rheological oscillatory and creep experiments. We observed that at higher temperatures the mechanism responsible for aging became faster thereby shifting the evolution of elastic modulus to lower ages. Significantly, in the creep experiments, all the aging time and the temperature dependent strain data superposed to form a master curve. Possibility of such superposition suggests that the rheological behavior depends on the temperature and the aging time only through the relaxation processes and both the variables do not affect the distribution but only the average value of relaxation times. In addition, this procedure allows us to predict long time rheological behavior by carrying out short time tests at high temperatures and small ages.

# I. Introduction:

Many soft colloidal materials such as concentrated suspensions, emulsions, gels, foams, pastes, etc. do not reach equilibrium over experimental time scales due to insufficient mobility of the constituent particles.[1] These 'soft glassy materials' (SGM) have a tendency to relax towards a possible equilibrium state over a prolonged period of time, which is known as aging and occurs at increasingly sluggish rates.[2-5] Understanding aging in soft glassy materials is very important, as it determines the long term physical behavior of these materials in industrial applications. Temperature has a very significant effect on the physical state and the glassy dynamics of soft glassy materials as it affects the microscopic mobility and the interparticle interactions.[6, 7] Usually the increase in temperature is known to cause jamming-unjamming transition.[8, 9] Although a lot is known about effect of temperature on the physical state, not much is known about its effect on the aging dynamics of soft colloidal materials. In this article we have investigated this very phenomena by carrying out rheological oscillatory and creep experiments. We observed that aging in soft colloidal materials became fast at higher temperatures. In addition, we observe the superposition of all the aging time and temperature dependent creep data, thereby enabling prediction of the long time rheological behavior of these materials by carrying out the short time tests.

Although understanding of the glassy dynamics in soft glassy materials is in its infancy, extensive literature is available on the glassy dynamics of molecular glasses.[10-13] In molecular glass formers, rapid decrease in temperature below the glass transition temperature causes volume contraction. The corresponding volume is however larger than the volume of the crystalline state at that temperature and pressure.[11, 14] In an isothermal annealing process, aging in molecular glasses is accompanied by volume recovery towards its equilibrium value as a function of time.[14] Kovacs[15] studied aging in various polymeric glasses and observed that isothermal volume recovery became faster as temperature approached the glass transition temperature. The horizontal shifting of such isothermal volume contraction curves yielded a superposition with corresponding shift factor showing Williams-Landel-Ferry (WLF) dependence on the temperature.[15, 16] In seminal contributions, Struik[12] and more recently O'Connell and McKenna[17] carried out time-aging time superposition for variety of polymers at different temperatures and obtained the dependence of shift rate $\mu$ ($=d\ln\tau/d\ln t_w$), which represents evolution dominating of relaxation time $\tau$ as a function aging time $t_w$, on temperature. It was observed that, for temperatures significantly below the glass transition temperature ($T_g$), $\mu$ either increased or remained constant with increase in temperature. However, as temperature approached $T_g$ from below, $\mu$ decreased so that the aging time dependence of the relaxation time faded away at $T_g$.[12, 17] Similar behavior of $\mu$ has also been reported for spin glasses.[18]

While aging in molecular glasses involves relaxation of the free volume, soft glassy materials experience evolution of the structure to a lower potential energy state,[19] which enhances their elasticity as a function of aging time.[20, 21] The evolution of the elasticity is usually monitored in a rheological experiment by applying small amplitude oscillatory shear. Krishnamoorti and coworkers[22] were the first to study the aging behavior of polymer clay nanocomposite melt, which shows a typical soft glassy behavior, as a function of temperature. They observed that isothermal evolution of elastic modulus with aging time shifted to smaller aging times for the experiments carried out at higher temperatures.[22] Similar observation has also been reported for other polymer clay nanocomposite systems,[23] aqueous clay suspensions,[24, 25] and other soft glassy materials.[24] Although all these reported observations for soft glassy materials are similar to the observation[15] of accelerated aging at higher temperatures for molecular glasses, the fundamental difference in characteristic features of aging in molecular and soft colloidal glasses demand a detailed investigation and a comprehensive analysis of the observed phenomenon.

## II Materials and Experimental Procedure

In this study we have used two systems. First system is an aqueous suspension of 3.2 wt. % laponite (LRD) while the second system is an aqueous suspension of 5 wt. % laponite with 0.5 wt. % Polyethylene oxide (PEO) having mean molar mass = 200 gm/mol (LRD-PEO). The second suspension system is expected to be in the nematic phase in which PEO is introduced to reduce the viscosity.[26] Laponite RD used in this study was procured from Southern Clay Products, Inc. Laponite was dried for 4 hours at 120 °C before mixing with water having a predetermined concentration of PEO (Loba Chemie) and pH 10. The basic pH was maintained by the addition of NaOH to provide chemical stability to the suspension. The suspension was stirred vigorously for 1.5 hours and left undisturbed for 2 weeks in a sealed polypropylene bottle. Soon after the preparation, aqueous suspensions of laponite with concentration above 1 vol. % (~2.6 wt. %) is known to form non-ergodic soft solids that withstand its weight.[27] In this work, we have used a stress controlled rheometer, AR 1000 (Couette geometry, bob diameter 28 mm with gap 1 mm). For an each test, shear cell was filled up with a new sample and allowed for the thermal equilibrium to be attained. The sample was then subjected to an oscillatory deformation with stress amplitude of 70 Pa and frequency of 0.1 Hz to carry out the shear melting. The shear melting experiment was stopped after attaining a plateau of low viscosity, from which the aging time was measured. The sample was allowed to age until a desired aging time was reached. Subsequently in a creep experiment, a constant stress of 5 Pa was applied to LRD while 15 Pa was applied LRD-PEO. Both the samples were very viscous and the respective values of stresses were chosen based on the criterion of less scatter in the strain response by carrying out experiments over a range of stresses. These creep stress values were within the linear response regime of the respective systems. To avoid evaporation of water or $CO_2$ contamination of the sample, the free surface of the suspension was covered with a thin layer of low viscosity silicon oil throughout the experiment.

## III. Results and Discussion

In an aging experiments, evolution of complex viscosity was monitored by applying an oscillatory shear stress having amplitude 0.1 Pa and frequency 0.1 Hz. It should be noted that, since the beginning of the experiment the elastic modulus ($G'$) was greater than the viscous modulus ($G''$) by at least one order of magnitude for both the systems. For the aging suspensions of laponite it is usually observed that after $G'$ crosses over $G''$, evolution of $G'$ is independent of frequency. This ensures system to be in the non-ergodic state.[19, 20] Inset in fig. 1 shows evolution of complex viscosity of LRD at four temperatures. It can be seen that with increase in temperature, an evolution of complex viscosity got shifted to a lower age. The horizontal shifting of the temperature dependent evolution curves lead to the superposition shown in Fig. 1. The corresponding shift factor $a(T)$ for both the systems is plotted against $1/T$ in fig. 2. It can be seen that the Arrhenius dependence (exponential relationship) fits the variation of shift factor as a function $1/T$ very well. However, since the experimentally explored range of $1/T$ is very limited, an Arrhenius dependence may not be the only possible relationship and the other functional forms may also fit the mentioned data.

We stopped the aging experiments at desired aging (waiting) times and carried out the creep experiments by applying a constant shear stress. Inset in fig. 3 shows a typical aging time dependent creep behavior, wherein strain showed weaker enhancement for the experiments carried out at the larger aging times. Joshi and Reddy[28] suggested that creep data obtained at various aging times demonstrates *creep time-aging time superposition*[12, 21] when product of transient creep compliance ($J(t_w + t) = \gamma(t_w + t)/\sigma_0$) and zero time modulus ($G(t_w)$) is plotted against $t/t_w^\mu$ in the limit of $t \ll t_w$.§ Here $t$ is creep time, $t_w$ is aging time and $\mu$ ($= d\ln\tau / d\ln t_w$) is a shift rate, which represents rate of evolution of relaxation time

---

§ Time-aging time superposition may not be always possible in soft materials, particularly when the $\alpha$ relaxation and the $\beta$ relaxation are not widely separated [5.]

arXiv:0907.0911

with respect to aging time. In fig. 3 we show the corresponding superposition for the age dependent creep data obtained at 30°C for $\mu = 0.45$. For both the systems explored in this work, we carried out such superposition procedure at four temperatures. In fig. 4, we have plotted the corresponding variation of $\mu$ for both the systems, which showed a linear dependence on $1/T$. A decrease in $\mu$ with an increase in temperature indicated that the rate of evolution of characteristic time scale ($d\ln\tau/d\ln t_w$) was slower at higher temperatures.

The observation that accelerated aging dynamics at higher temperature was accompanied by decrease in $\mu$ (slower evolution of relaxation time) appears to be counter intuitive. In the following discussion we propose a mathematical framework to analyze this phenomenon. In a nonergodic state, the system can be represented by particles trapped in potential energy wells created by their neighbors. In such a state, mere thermal motion is not sufficient to cause diffusion out of a cage or a well.[19] However, even though such jamming limits the translational mobility of the arrested particles, they are free to undergo microscopic structural rearrangements within the cage. While undergoing the microscopic rearrangements, the particles prefer those configurations that have lower energy.[19] This causes particles to progressively sink in their energy wells with the aging time. Therefore, since the decrease in the energy of a particle with the aging time is by virtue of microscopic rearrangement dynamics, the depth of the energy well or the barrier height ($E$) should depend on the age normalized by the timescale of the microscopic motion, $(t_w/\tau_m)$; leading to:

$$E = E(t_w/\tau_m). \tag{1}$$

Generally, depending upon the nature of a corresponding cage and by virtue of a location and an orientation, particles that are trapped in the energy wells have a distribution of potential energy barriers. However, for simplicity, we will consider a situation where all the wells have same barrier height $E$. If $b$ is the characteristic

arXiv:0907.0911

length-scale of the microstructure, elastic modulus of the system can be related to the barrier height $E$ as,[14]

$$G' = E(t_w/\tau_m)/b^3 .  \qquad (2)$$

If we assume that characteristic length-scale of the microstructure does not depend on aging time, eq. 2 suggests that $G'$ also depends on $t_w/\tau_m$. In fig. 1, we observed that evolution of storage modulus (fig. 1 reports evolution of complex viscosity, however since $G'' \ll G'$, $G' \approx \omega|\eta^*|$) got shifted to a lower age at higher temperatures. Therefore, in order to obtain the superposition by horizontal shifting, the shift factor reported in fig. 1 should scale as, $a(T) \sim 1/\tau_m$. Microscopic timescale $\tau_m$ characterizes motion of a particle within the cage and is expected to become fast with increase in the temperature. As shown in fig. 2, Arrhenius relationship fits dependence of $a(T)$ on temperature very well. Although the other functional forms are also possible, we assume the Arrhenius relationship in the present analysis leading to a following expression for the microscopic time scale:

$$a(T) \sim 1/\tau_m = (1/\tau_{m0})\exp(-\bar{U}/k_B T), \qquad (3)$$

where $\bar{U}$ is the activation barrier associated with the microscopic rearrangement of the particle.

The phenomenon of aging time dependent enhancement in elasticity of soft glassy materials is similar to aging time dependent volume recovery in molecular glasses observed by Kovacs,[15] who observed WLF type of temperature dependence for the shift factor. The WLF type of dependence of the shift factor can be derived by considering diffusion of the free volume to be responsible for the volume relaxation.[29] On the other hand, for soft materials, we propose that the microscopic timescale for structural rearrangements sets the overall timescale for the aging process in soft glasses, which becomes fast with increase in the temperature. The present work is the first study that investigates the temperature dependence of the shift factor in the aging experiments of the soft glassy materials.

arXiv:0907.0911

For the glassy materials that do not obey time translational invariance, the evolution of the characteristic relaxation time (cage diffusion time) often shows a power law dependence on the aging time given by,[12, 19]

$$\tau = \tau_m^{1-\mu} t_w^{\mu}. \tag{4}$$

On the other hand, the same time scale is related to the barrier height $E$ by an Arrhenius relationship:

$$\tau = \tau_m \exp(E/k_B T). \tag{5}$$

In the present experiments we have used very small values of stresses, and therefore assume that the energy[19] ($\frac{1}{2}k\gamma^2$) gained by the particle due to strain ($\gamma$) induced in the system is too small compared to the barrier height $E$. Eqs. 4 and 5 lead to expression of $\mu$ given by, $\mu = d(E/k_B T)/d\ln t_w$. However, in order to have a constant value of $\mu$ over the range of explored waiting times, the elastic modulus must show a logarithmic dependence on the aging time given by:

$$G' = E/b^3 = G_0 \ln(t_w/\tau_m). \tag{6}$$

It can be seen in fig. 1 that above 3000 s, the minimum aging time at which creep experiments were performed, evolution of complex viscosity (Since $G'' \ll G'$, $\eta^* \cong G'/\omega$) can indeed be approximated by a logarithmic dependence on aging time. Eqs. 5 and 6 lead to an expression for the characteristic relaxation time given by,

$$\ln \tau = \left[b^3 G_0 / k_B T\right] \ln t_w + \left[1 - (b^3 G_0 / k_B T)\right] \ln \tau_m \tag{7}$$

which has the same functional form as eq. 4, and hence leads to an expression for the shift rate $\mu$ as,

$$\mu = b^3 G_0 / k_B T. \tag{8}$$

Remarkably, this relationship qualitatively explains the linear dependence of $\mu$ on $1/T$ for both the systems as shown in fig. 4. Therefore, even though the evolution of elastic modulus shifts to smaller ages with the increase in temperature, rate of increase in characteristic time scale $\mu$ ($=d\ln\tau/d\ln t_w$) actually decreases with temperature. This difference is essentially due to the fact that the elastic modulus

arXiv:0907.0911

is proportional to the barrier height $E$, while the relaxation time depends on $E/k_B T$ through the Arrhenius relationship.

The respective creep time-aging time superpositions, as shown in fig. 3, at various temperatures showed the same curvature leading to a natural possibility to obtain a *creep time-aging time-temperature* superposition. In fig. 5, we have plotted the master superposition for both the systems by horizontally shifting various isothermal creep time-aging time superpositions. Each master superposition contains 16 creep curves obtained at different aging times and temperatures. It can be seen that the superposition belonging LRD-PEO (fig. 5a) showed significant scatter compared to that of LRD (fig 5b). It is usually observed for a creep flow of laponite suspension that scatter in the strain response increases with increase in the concentration of laponite. The effect becomes more pronounced with decrease in the temperatures as well as the stress.[25, 28] Possibility of such superposition implies that aging time and temperature do not affect the distribution of relaxation times but affect only its average value. As shown in insets of figures 5 (a) and 5(b), the logarithm of corresponding horizontal shift factor, $\ln c(T)$, showed linear dependence on $1/T$ with a negative slope. If we assume that the short time creep behavior depends on the age and the temperature of the sample only through its dependence on the relaxation time, the horizontal shift factor $c(T)$ shown in fig. 5 can be related to creep time and aging time by:

$$c(T) \frac{t}{t_w^\mu} = \frac{t}{\tau} = \frac{t}{\tau_m^{1-\mu} t_w^\mu} \tag{9}$$

This leads to a power law relationship between $c(T)$ and $\tau_m$ given by $c(T) \sim \tau_m^{\mu-1}$, which combined with expression for $\tau_m$ described by eq. 3 leads to,

$$\ln c(T) \sim (\mu-1)\left(\ln \tau_{m0} + \bar{U}/k_B T\right) \sim \frac{\left(b^3 G_0 \ln \tau_{m0} - \bar{U}\right)}{k_B T} + \frac{b^3 G_0 \bar{U}}{(k_B T)^2} - \ln \tau_{m0}. \tag{10}$$

The first term on the right hand side of Eq. 10 leads to a linear dependence of $\ln c(T)$ on $1/T$ while the second term leads to a quadratic dependence on $1/T$. The

arXiv:0907.0911

observed dependence of $\ln c(T)$ shown in fig. 5 implies that the magnitude of the first term on the right hand side of Eq. 10 is larger than the second term and opposite in sign. It should be noted that the scaling analysis presented above involves only a simple scenario of monodispersed barrier height distribution. Interestingly the analysis nicely captures the temperature dependences of various parameters.

Overall, this article leads to some significant results which give unique insight into the aging in laponite suspensions. Interestingly aging in laponite suspensions shows some striking similarities with that of in polymeric glasses. The first one is the faster aging dynamics at high temperature observed in both the systems. We believe that the observed dependence on temperature is a signature of the mechanism for aging which is at play. It is proposed in the literature that application of stress or the deformation field enhances the effective temperature of the soft glassy materials.[19] Therefore, it would be interesting to study effect of the same on the aging dynamics. Another prominent result is the observation of decrease in $\mu$ with increase in temperature. Significantly polymeric glasses also demonstrate decrease in $\mu$ as the glass transition temperature ($T_g$) is approached from below,[12, 17] because at $T_g$ material becomes ergodic and does not show the time dependence. In the present system, however, the observed temperature dependence of $\mu$ originates from Arrhenius dependence of the dominating relaxation time (eq. 5). In addition, we believe that the observed linear dependence of $\mu$ on $1/T$ in the present system is possible only in the limit of small strains (or stresses). This is because at sufficiently higher stresses, the potential energy enhancement of the particle $\frac{1}{2}k\gamma^2$ due to induced strain $\gamma$ may lead to partial yielding of the sample by reducing the relative barrier height.[19] Such partial yielding is known to lower the cage diffusion time scales and also may affect the temperature dependence. Finally, we demonstrated the *creep time-aging time-temperature superposition*, suggesting that the rheological behavior depends on the temperature and the aging time only through the relaxation processes and both the variables do not affect the spectrum of relaxation times but only the average value. The procedure developed in the present paper also enables estimation of long time rheological behavior of soft materials at low temperatures and high age by carrying out short time tests at high temperatures and small age. However it should be noted that the present superposition is limited to the region $J(t+t_w)G(t_w) = O(1)$. This is due to the creep time data considered in the superposition is in the limit $t << t_w$. In order to consider the larger creep time data, one must consider the aging of the sample as it undergoes creep. We have also developed a simple scaling model that qualitatively captures the observed temperature dependence of $\mu$ and that of horizontal shift factor $c(T)$, required to obtain the *creep time-aging time-temperature superposition*. The present work therefore leads to important insights into the aging phenomena of soft glassy materials and demonstrates various similarities in aging behavior of soft glassy materials and molecular glasses.

## IV. Conclusions:

In this article we study effect of temperature on the aging behavior of soft glassy materials by carrying out creep and oscillatory experiments. We observed that the evolution of elastic modulus, which accompanies aging, got shifted to the lower age at higher temperatures, suggesting that the mechanism responsible for aging becomes faster at higher temperatures. The subsequent creep time-aging time superposition yielded rate of evolution of characteristic relaxation time on age $\mu$ ($=d\ln\tau/d\ln t_w$), which however, decreased with increase in temperature. We have developed a simple scaling model, which shows that the observed behavior can be qualitatively explained by expressing elastic modulus to be directly proportional to barrier height of the energy well ($E$) and the Arrhenius dependence of relaxation time. Finally the temperature dependent creep time-aging time superpositions lead to *creep time-aging time-temperature superposition*. This suggests that the rheological behavior of the soft glassy materials only implicitly depends on temperature and aging time through its dependence on relaxation time-scales. In addition, such superposition enables estimation of long time rheological behavior at low temperatures and high ages by carrying out short time tests at high temperatures and small ages.

**Acknowledgement**: We would like to dedicate this article to Dr. B. D. Kulkarni on the occasion of his 60th birthday. We thank Prof. P. Coussot for his comments on the manuscript. This work was supported by Department of Science and Technology, Government of India through IRHPA scheme.

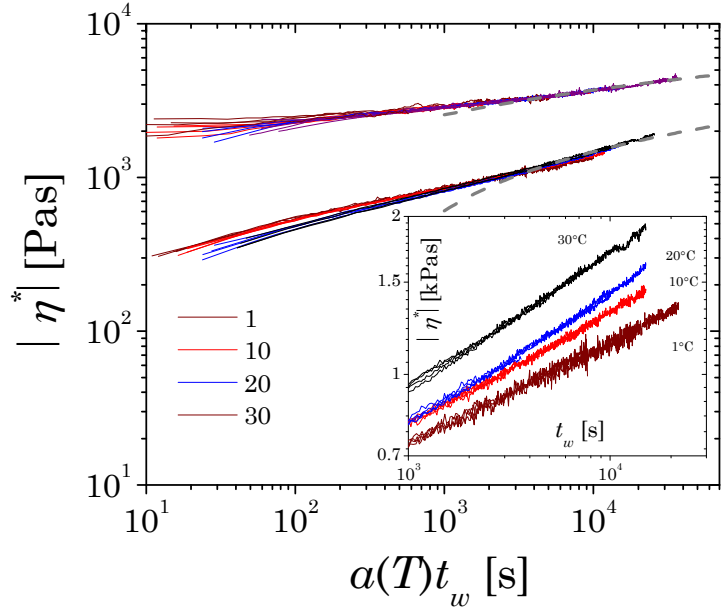

**Figure 1.** Master curves describing evolution of complex viscosity with age (Since $G'' \ll G'$, $\eta^* \cong G'/\omega$). Top master-curve belongs to LRD-PEO while bottom master-curve belongs to LRD. Superposition is obtained by horizontally shifting the data at various temperatures to the data at 20 °C. Dashed lines show logarithmic fits to the evolution data. Inset shows evolution of complex viscosity of LRD at four temperatures.

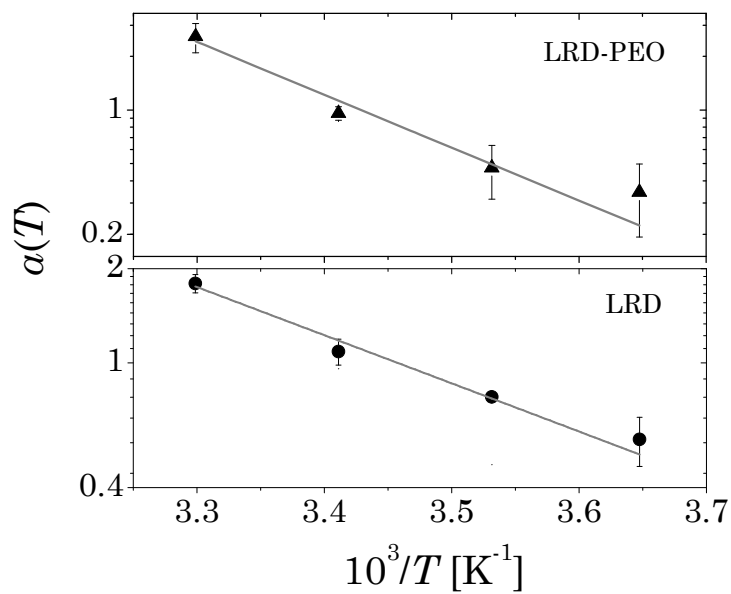

**Figure 2.** Temperature dependent shift factor $a(T)$ shown in fig. 1 is plotted against $1/T$. The line represents Arrhenius dependence.

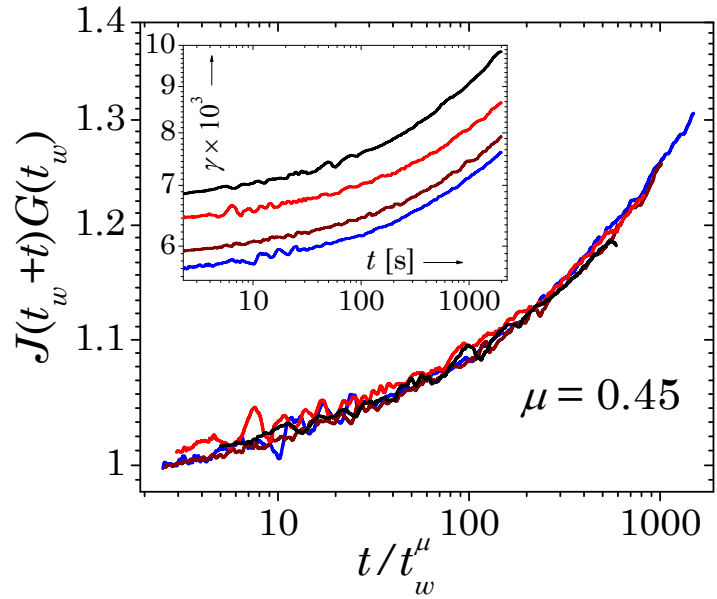

**Figure 3.** Inset shows creep curves obtained at various aging times ($t_w$) at 30°C plotted against creep time ($t$) for LRD-PEO (From top to bottom $t_w$ = 3000, 7000, 10000 and 15000 s). Main figure shows creep time-aging time superposition wherein normalized creep compliance is plotted against $t/t_w^\mu$ for the creep curves shown in the inset. It can be seen that the superposition is obtained for $\mu$ =0.45.

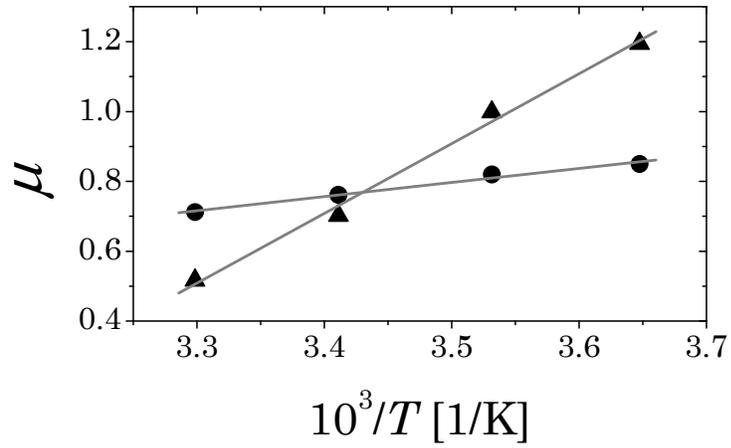

**Figure 4.** Shift rate $\mu$ defined as $\mu$ $(=d\ln\tau/d\ln t_w)$ plotted against reciprocal of temperature. Filled triangles represent LRD-PEO while filled circles represent LRD.

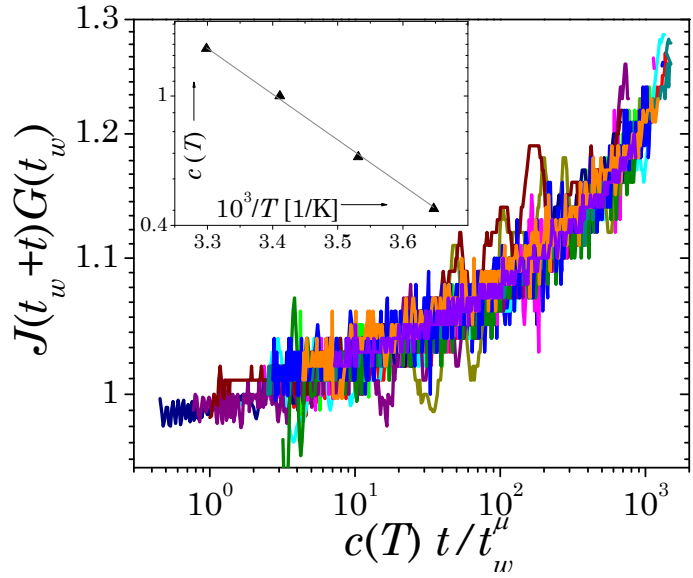

(a)

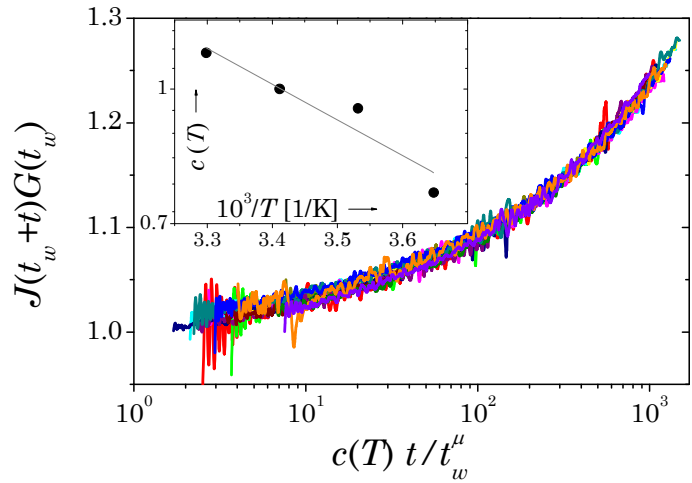

(b)

**Figure 5.** Creep time – aging time – temperature superposition for (a) LRD-PEO and (b) LRD. For both the systems individual creep time – aging time superpositions at various temperatures were shifted horizontally by using shift factor $c(T)$, whose dependence on temperature is given in the insets.